\pdfoutput=1
 \documentclass[aps,prl,twocolumn,groupedaddress]{revtex4-1}
 \usepackage{calrsfs}
 \usepackage{mathrsfs}
 \usepackage{graphicx}
 \usepackage{dcolumn}
 \usepackage{bm}
 \usepackage{float}
 
 \usepackage{hyperref}
 \usepackage{xcolor}
 \hypersetup{colorlinks=true,linkcolor=red}
 \usepackage{color}
 \usepackage{graphicx}
 \usepackage{fancyhdr}
 \usepackage{url}
 \usepackage{amssymb,amsfonts,amsmath,amsbsy,bm,t1enc,lipsum,latexsym}
\begin{document}
\title{Bringing short-lived dissipative Kerr soliton states in microresonators into a steady state}

\author{Victor Brasch}
\affiliation{{\'E}cole Polytechnique F{\'e}d{\'e}rale de Lausanne (EPFL), Institute of Physics, CH-1015 Lausanne, Switzerland}
\author{Michael Geiselmann}
\affiliation{{\'E}cole Polytechnique F{\'e}d{\'e}rale de Lausanne (EPFL), Institute of Physics, CH-1015 Lausanne, Switzerland}
\author{Martin H. P. Pfeiffer}
\affiliation{{\'E}cole Polytechnique F{\'e}d{\'e}rale de Lausanne (EPFL), Institute of Physics, CH-1015 Lausanne, Switzerland}
\author{Tobias J. Kippenberg}
\email[E-mail: ]{tobias.kippenberg@epfl.ch}
\affiliation{{\'E}cole Polytechnique F{\'e}d{\'e}rale de Lausanne (EPFL), Institute of Physics, CH-1015 Lausanne, Switzerland}

\begin{abstract}
	Dissipative Kerr solitons have recently been generated in optical microresonators, enabling ultrashort optical pulses at microwave repetition rates, that constitute coherent and numerically predictable Kerr frequency combs. However, the seeding and excitation of the temporal solitons is associated with changes in the intracavity power, that can lead to large thermal resonance shifts during the excitation process and render the soliton states in most commonly used resonator platforms short lived. Here we describe a ``power kicking'' method to overcome this instability by modulating the power of the pump laser. A fast modulation triggers the soliton formation, while a slow adjustment of the power compensates the thermal effect during the excitation laser scan. With this method also initially very short-lived ($\mathcal{O} \sim 100 \mathrm{ns}$) soliton states , as encountered in $\mathrm{Si}_3\mathrm{N}_4$ integrated photonic microresonators, can be brought into a steady state in contrast to techniques reported earlier which relied on an adjustment of the laser scan speed only. Once the soliton state is in a steady state it can persist for hours and is thermally self-locked. 
\end{abstract}

	\maketitle

	Optical frequency combs provide a set of equidistant laser lines in the optical domain \cite{Udem2002,Cundiff2003} and have revolutionized time keeping, frequency metrology, as well as spectroscopy in the last decade.  \cite{Newbury2011}. Discovered in 2007, Kerr frequency combs \cite{DelHaye2007, Kippenberg2011}  allow for the generation of optical frequency combs from a single continuous wave (CW) pump laser via the Kerr nonlinearity inside a microresonator. Such Kerr combs offer access to repetition rates in the microwave domain, broadband spectra and chipscale integration, and have been employed in a growing number of applications including terabit communication \cite{Pfeifle2014} or optical atomic clocks \cite{Papp2014}.
	Recently a new operating regime has been observed  in crystalline microresonators \cite{Herr2013} in which Kerr frequency combs give rise to the spontaneous formation of dissipative temporal solitons \cite{Leo2010, Grelu2012, Akhmediev2005}.
	Dissipative Kerr solitons (DKS) represent a pulsed waveform which maintains its shape indefinitely due to a double balance of the dispersion and the nonlinearity as well as the losses and the nonlinear parametric gain inside the microresonator \cite{Leo2010, Grelu2012}. The resulting stable waveform gives rise to a fully coherent optical frequency comb and mitigates the noise processes which have been shown to result in excess phase noise for earlier broadband Kerr frequency combs \cite{DelHaye2011, Herr2012}. For this reason bright DKS are the preferred operating regime for the generation of Kerr frequency combs and have been demonstrated in a variety of microresonator platforms, such as crystalline microresonators \cite{Herr2013, Liang2015}, silica disk microresonators \cite{Yi2015} and silicon nitride integrated microresonators \cite{Brasch2016, Wang2016, Joshi2016}. 
	However, due to thermal effects inside the microresonator related to the strong pump laser it can be challenging to bring the soliton states in a steady state. Here we describe a method to overcome this problem also for very short-lived soliton states by modulating the pump power in order to obtain the soliton state at a precise point in time and to bring them to a steady state. The technique described here has been developed for soliton states in integrated microresonators made from silicon nitride \cite{Levy2010,Moss2013} and has been first used in \cite{Brasch2016} for soliton generation in $\mathrm{Si}_3\mathrm{N}_4$ microresonators but has also been adapted for silica disk microresonators \cite{Yi2015, Yi2016}.
	
\begin{figure}[H]
		\centering \includegraphics[width=1\columnwidth]{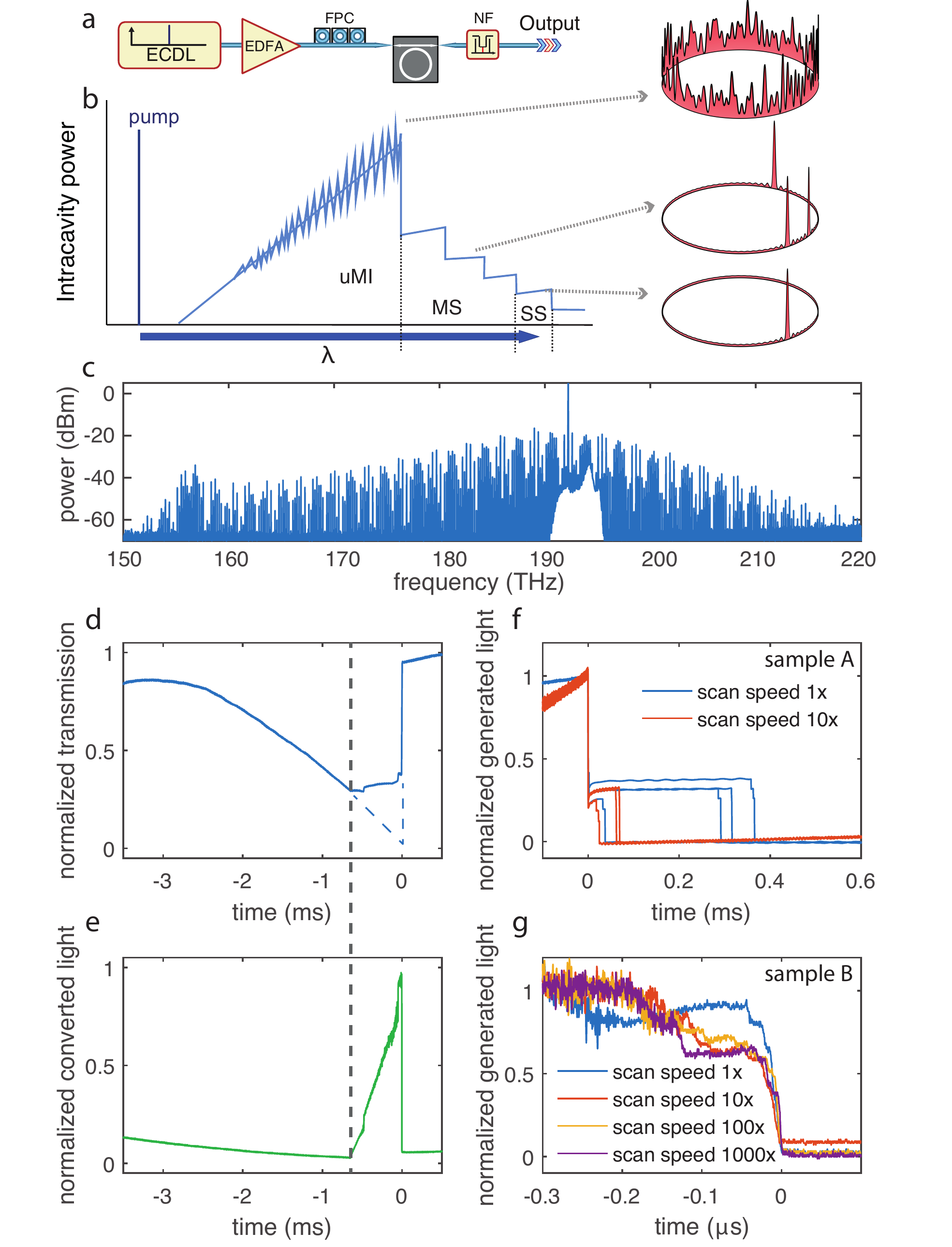}
		\caption{\textbf{Soliton step observation in $\mathrm{Si}_3\mathrm{N}_4$ microresonators. } (\textbf{a}) Setup used to observe soliton steps. A narrow-bandwidth filter (NF) is implemented to filter out the strong pump light. ECDL: external cavity laser diode; EDFA: erbium-doped fiber amplifier; FPC: fiber polarization control. (\textbf{b}) Schematic of the intracavity power during a laser scan over a resonance. Tuning a cw laser into a resonance produces first the unstable modulation	 instability (uMI)  before reaching the soliton regime. Each step represents a different number of solitons propagating in the cavity. (MS: multi-soliton, SS: single soliton) (\textbf{c}) Optical spectrum of a four-soliton state, repetition rate is 190\,GHz. (\textbf{d}) The thermal triangle transmission profile that is typical for a scan of a strong pump laser over a microresonator resonance from higher to lower frequencies. The shape is caused by optical and thermal nonlinearities. The triangle flattens out once the parametric threshold is reached (dashed vertical line). (\textbf{e}) The converted light is the light that is converted from the pump laser to the frequency comb by the Kerr nonlinearity. The pump laser is suppressed with a narrow-bandwidth filter (shown in \textbf{a}). The soliton steps occur at the sharp trailing edge of the converted light trace when the state inside the microresonator changes from the  uMI to the soliton state. (\textbf{f}) For this sample the soliton steps have a duration of the order of $\mathcal{O}(0.1)\,\mathrm{ms}$ and the step duration changes as the pump laser scan speed is changed from 300\,GHz/s to 3\,THz/s. (\textbf{g}) For another sample the steps are only around 0.1\,$\mu$s long and although the scan speed is changed by three orders of magnitude from 3\,GHz/s to 3\,THz/s, the step pattern remains very similar. All traces are from integrated silicon nitride microresonators: sample A is fabricated with the photonic Damascene Process \cite{Pfeiffer2016} (nominal cross-section 0.85x1.5$\mu$m) and B with the subtractive process \cite{Brasch2016} (nominal cross-section 0.8x1.77$\mu$m).}
		\label{fig:steps}
	\end{figure}

	DKS states in microresonators are typically accessed from the unstable modulation instability (uMI) regime by scanning the pump laser frequency from higher to lower frequencies over a resonance of the microresonator \cite{Herr2013} (Fig. \ref{fig:steps}a,b). Because of nonlinear optical and thermal effects of the strong pump laser the resonance is pulled along with the laser and the lineshape of the resonance then takes a triangular shape instead of the otherwise typical Lorentzian one \cite{Carmon2004} (Fig.\ref{fig:steps}d). 
	The end of the thermal triangle is marked by the point where the nonlinear shift is maximum which occurs when the pump laser reaches the zero-detuning point with respect to the shifted resonance. At this point the pump laser looses the resonance and the transmission  transitions back to its value far detuned from the resonance, on the timescale of the thermal effect. The converted light, i.e. the light that is converted from the pump to new frequencies via the Kerr nonlinearity inside the microresonator, shows the inverse behavior once the threshold power is reached (at around --0.6\,ms in Fig.\ref{fig:steps}e). These sharp transitions of the thermal triangles at the zero-detuning point coincide with the transition from the uMI state to the soliton state \cite{Herr2013} (Fig. \ref{fig:steps}b). One sign of the existence of DKS states inside microresonators are step patterns that can be 	observed when scanning the pump laser past this sharp transition \cite{Herr2013} (Fig.\ref{fig:steps}f and g for examples from silicon nitride waveguide microresonators). For the duration of these steps the Kerr frequency comb is brought into a soliton state. For consecutive laser scans the exact step pattern typically varies as the transitions from states with higher number of solitons to a lower number of solitons that result in the individual steps occur at different times. 
	
	For most resonators the soliton states within these steps are not stable such that they will not persist when the pump laser scan is simply stopped within the steps, or the steps are so short in duration that stopping the laser scan within the steps is technically challenging. The reason for this is the thermal effect, caused by residual absorption of the pump light in the microresonator \cite{Illchenko1992, Carmon2004, Xue2016}. In the uMI state, the intracavity power is typically substantially higher than in the soliton state, which can be seen for example in Fig.\ref{fig:steps}f in the large first step occurring at time 0\,ms. This designates the transition from the uMI state to the first multiple soliton state. The change in the coupled resonator power consequently leads to a larger thermal shift of the resonance for the uMI state than in the steady state soliton state. If the pump laser scan is stopped within the soliton step just after the transition from the uMI state, the substantially lower intracavity power in the soliton state leads to a cooling of the resonance with a resulting smaller thermal shift during the forward laser scan. If the cavity cools an amount that is appreciable in magnitude compared to the detuning range over which the soliton state can be maintained (the so called soliton existence range \cite{Karpov2016}) the cooling of the resonator can lead to an extinction of the solitons inside the cavity. Therefore it is necessary to compensate this difference in thermal shifts.
	
	The difference in the impact of the \emph{magnitude} of the thermal effect is experimentally shown in Fig.\ref{fig:steps}f and g for $\mathrm{Si}_3\mathrm{N}_4$ microresonator of similar dimensions (and thus practically identical thermal time constant), but with different amount of thermal absorption that may be related to the two different fabrication processes.
	While for the sample A in Fig.\ref{fig:steps}f the soliton steps change in duration when the laser scan speed is changed (similar to the case for solitons in crystalline microresonators \cite{Herr2013}), the step pattern (i.e. the duration) remains very similar for the case of Fig.\ref{fig:steps}g (sample B). This behavior can be explained by the different amounts of thermal heating, and thus related to the different amounts of thermal shift that the cavity undergoes when cooling while tuning over the soliton existence range. For a small temperature increase, the amount of cavity cooling once transitioned from the uMI to the soliton state in terms of absolute frequency shift can be small compared to the soliton existence range. In this case the duration over which the soliton state is observed in a forward scan shortens with increasing scan speed (as the soliton existence range remains unaltered for different laser scan speeds). This behavior is observed in sample A. In contrast, the sample B shows a duration of the step which is \emph{independent} on the scan speed. This can be explained by noting that a larger absorption (and therefore higher temperature increase) will cause a larger cooling induced shift once it has transitioned from the uMI to the soliton regime. Once the thermal cooling shift becomes comparable to the soliton existence range, the scan speed dependence will diminish, and can become independent of the scan speed. This explanation is also in agreement with the observation that the steps can become power dependent and are observed to lengthen with lowering of the pump power (which reduces the amount of frequency shift the cavity cools during the scan through the soliton existence range).

	\begin{figure}[h!]
		\centering \includegraphics[width=1\columnwidth]{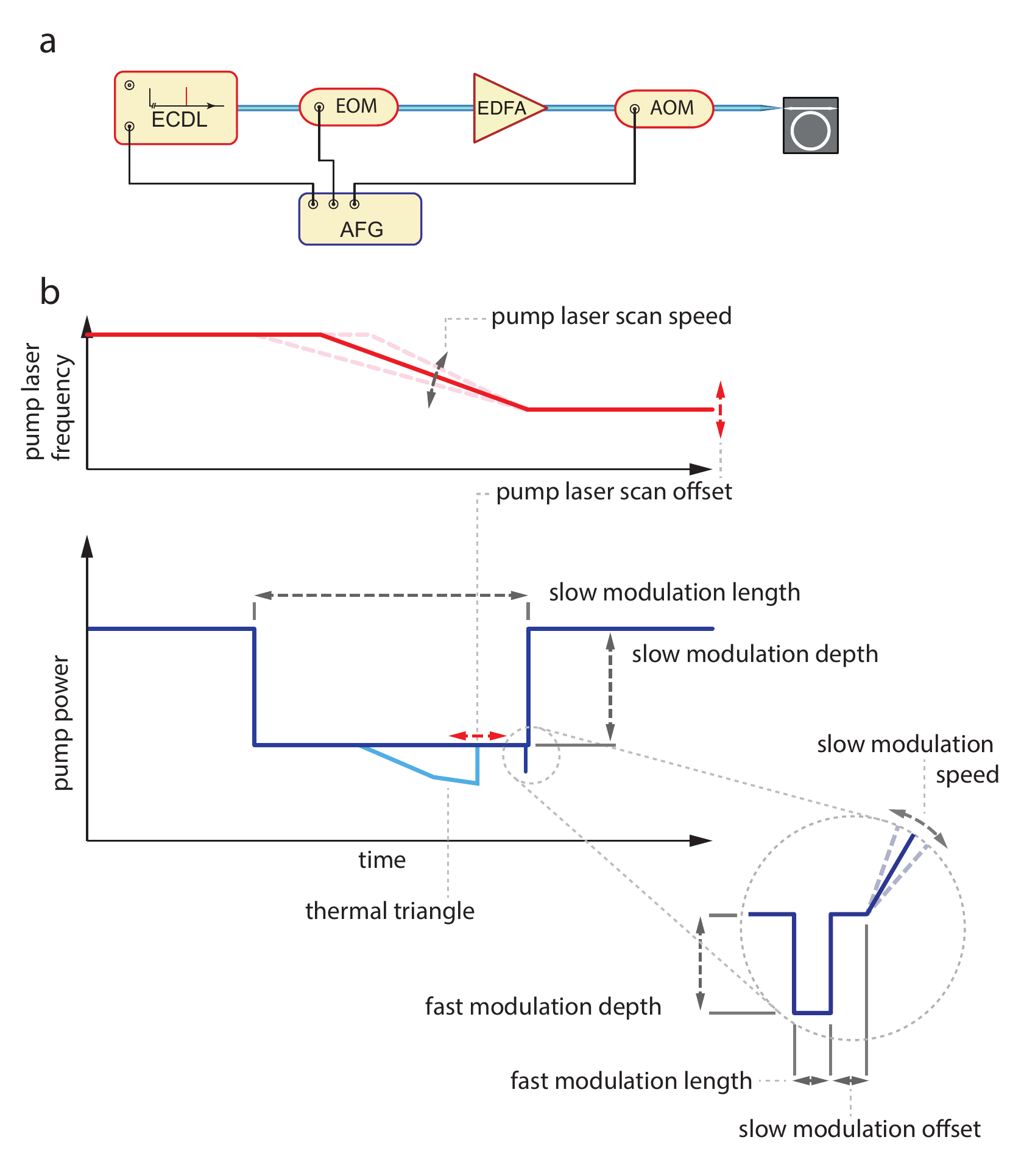}
		\caption{ \textbf{Setup and schematic showing the different parameters for the modulation
				approach.} (\textbf{a}) Setup used to bring the solitons to a steady state, including two modulators to adjust the pump power. ECDL: external cavity diode laser; EOM: Mach-Zehnder
			electro-optic amplitude modulator; EDFA: erbium-doped fiber amplifier;
			AOM: acousto-optic modulator; AFG: arbitrary function generator. The
			EOM and the AOM allow for a modulation of the pump power on a fast
			(EOM, AC-only, few nanoseconds) and slower (AOM, DC, above 100 ns)
			time scale. The AFG synchronizes the laser scan with the fast and
			slow pump power modulation in order to stabilize the soliton state. (\textbf{b}) When the solitons steps behave as shown in Fig.\ref{fig:steps}f, an adjusted pump laser scan (top) can already result in stable soliton states. The pump power increase (the ``slow modulation'') can help stabilizing the soliton states as shown in the middle. At the bottom right, the ``fast modulation'' of the pump power is used to obtain soliton states at a precise point in time is shown. This fast modulation is required because for very short soliton steps such as the ones shown in Fig.\ref{fig:steps}g a reliable timing of the pump power increase to occur within the soliton steps is otherwise not possible.}
		\label{fig:modulationScheme} 
	\end{figure}

	In order to bring the short-lived soliton states within the steps to a steady state, several measures can be taken that are described in the following. The first step is to adjust the scan speed of the pump laser and to attempt to stop the laser scan within the soliton steps. If the steps behave as shown in Fig.\ref{fig:steps}f, then there is a good probability that with the right laser scan speed the thermal instability can be circumvented \cite{Herr2013}. This can happen because for faster laser scans on a timescale of the thermal effect inside the microresonator, the microresonator does not reach thermal equilibrium during the laser scan. As a consequence by choosing the right laser scan speed the thermal shift within the uMI state can be limited to an amount close to the shift within the soliton state and therefore the transition from the uMI state into the soliton state will only result in a small change of the thermal shift which does not destabilize the soliton state. This method was initially used to obtain steady state soliton states in crystalline microresonators and is described in the SI of Ref.\cite{Herr2013} but has also successfully been implemented for silicon nitride microresonators \cite{Kordts2016, Pfeiffer2016, Karpov2016}. 
	
	For the case when adjusting the laser scan speed does not yield stable soliton states we have developed the pump power kicking modulation technique. If the steps are sufficiently long such that a reliable timing of a power modulation to occur within the steps is possible, it can help to introduce a pump power increase within the steps. In this scheme the increase in pump power leads to a higher absorption and therefore a stronger thermal effect. By choosing the pump power levels well it is possible to adjust the thermal shift of the pumped resonance within the soliton state with higher pump power to a similar level as for the uMI state with lower pump power. Such a pump power modulation can be implemented for example using an acousto-optic modulator (AOM) in the pump path. In this configuration the laser scan and the AOM for the slow power
	modulation are timed such that the AOM lowers the pump power before
	the laser tunes into the resonance and the AOM increases the pump
	power within the soliton steps at the end of the thermal triangle of the resonance. By how much exactly the AOM modulates the power has to be adjusted such that the soliton steps reach a steady state. On longer time scales the pump power can also be adjusted with an active feedback loop \cite{Yi2016}.

	\begin{figure}[h!]
		\centering \includegraphics[width=1\columnwidth]{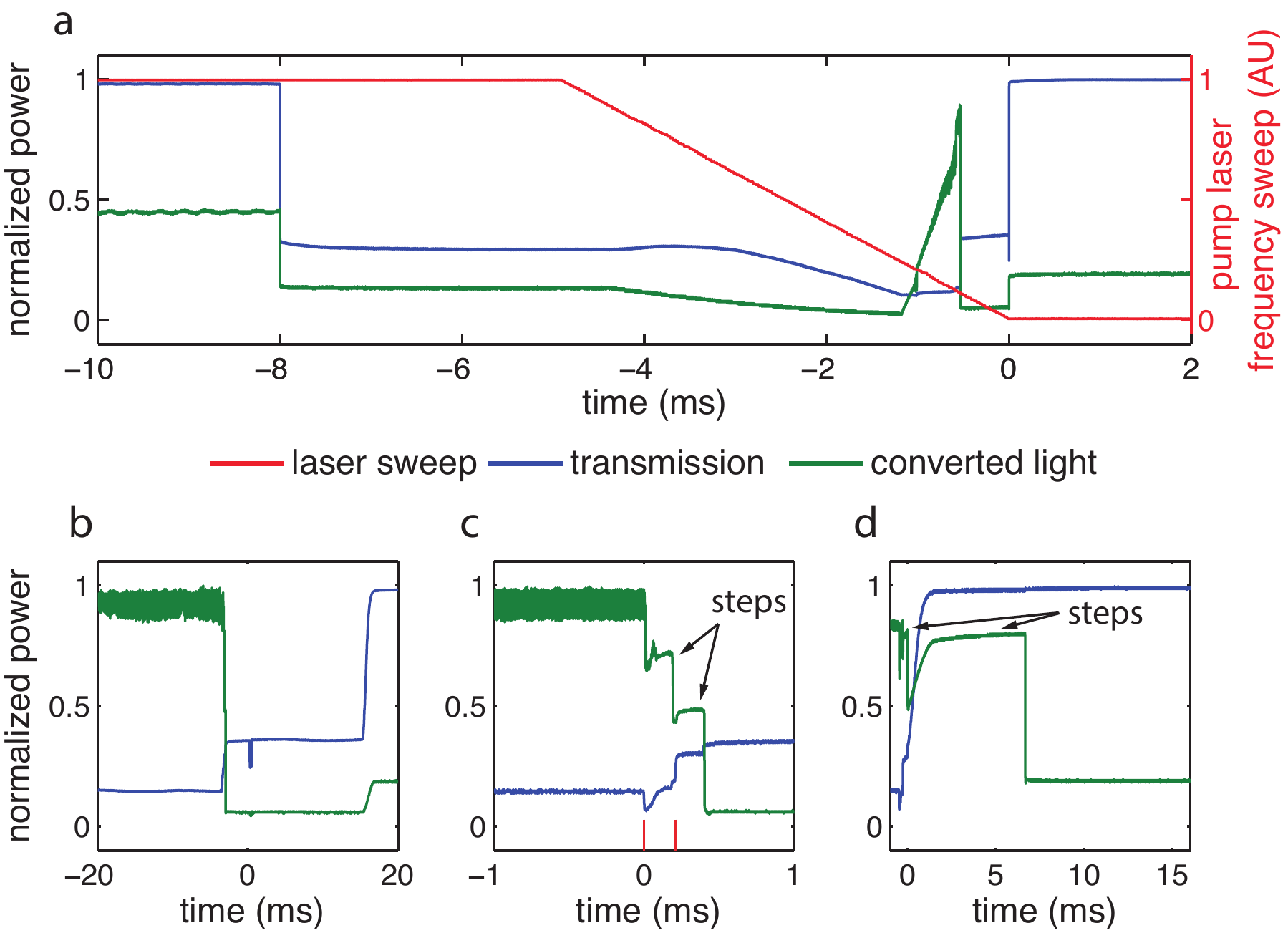}
		\caption{ \textbf{Modulation sequence for the stabilization of soliton steps.} (\textbf{a})The slow power modulation (clearly visible as a drop in transmission at
			\textendash 8\,ms and an increase at 0\,ms) on a larger time scale with
			respect to the laser scan (red) that sweeps over the resonance from
			higher to lower frequencies and stops just before 0 ms.
			Also shown is the converted light (green trace) which increases once
			the threshold is reached. (\textbf{b}) The initial timing of the fast
			modulation (small dip in the transmission at 0\,$\mu$s) with respect
			to the thermal triangle and the slow power modulation. (\textbf{c}) shows how
			the fast power modulation induces the soliton steps if set up properly. The
			fast modulation starts at 0\,$\mu$s and decreases the power for around
			200\,ns (marked with red lines). The slow increase of power is far away
			(as visible in b) and does not have any effect here. (\textbf{d}) Combined
			effect of the fast and the slow modulation when the slow increase is timed just
			after the fast modulation. Then the fast modulation induces the soliton steps at
			0\,$\mu$s after which the slow increase of pump power leads to
			much longer steps (visible in the green trace that stays up for around
			7\,$\mu$s). With further optimization these steps become a steady
			state and the resulting Kerr frequency combs can be passively stable for hours
			without any further measures.}
		\label{fig:powerKick} 
	\end{figure}
	
	If the soliton steps are very short as for example shown in Fig.\ref{fig:steps}g a reliable timing of a pump power increase to occur within the steps is not possible as the exact occurrence time of the steps typically varies by substantially more than the step duration and implementing a feedback loop with only 100\,ns delay or less is technically very challenging. Instead it is possible to obtain the soliton steps at a known time in a reliable way using a very fast pump power decrease which can be induced with for example an electro-optic amplitude modulator (EOM). This fast pump power drop can be seen as a way to effectively induce the zero-detuning transition of the pump laser with respect to the shifted resonance at the time of the pump power modulation as the lower pump power reduces the nonlinear shifts of the resonance and therefore also the detuning. This deterministic generation of soliton states at a particular point in time is crucial for the successful, steady state generation of soliton states with steps of only nanoseconds duration.
	The setup used in \citep{Brasch2016} is shown in Fig.\ref{fig:modulationScheme}a. It splits the power modulation into a very fast, ``AC-coupled'' part with an EOM before the EDFA and an AOM after the EDFA for slightly slower but DC modulation of the power. The optical AC-coupling of the EOM modulation is due to the EDFA, which runs with constant output power and acts therefore as a high-pass filter for optical power modulations. In principle the two pump power modulations (the quick drop and the increase to stabilize the soliton) can be implemented using the same intensity modulator if the modulator is fast enough and can handle the full pump power.
	
	In what follows we describe how to dial in the different modulation parameters that are shown in Fig.\ref{fig:modulationScheme}b such that steady-state solitons can be obtained for this approach with two power modulations. Initially the slow pump power modulation is dialed in such that the pump power is decreased before the pump laser tunes into resonance and the pump power is increased again to the previous level after the end of the thermal triangle. This can be seen in Fig. \ref{fig:powerKick}a
	in the transmission trace where the AOM lowers the pump power at --8\,ms
	and increases the pump power again at around 0\,ms.
	
	The EOM provides the fast power modulation that should result in soliton states at a defined point in time. Therefore timing of the
	EOM is such that it lowers the pump power further just before the
	AOM increases the pump power (Fig.\ref{fig:modulationScheme}b and
	\ref{fig:powerKick}b). For a start the separation between the EOM
	modulation and the AOM power increase (slow modulation offset) should
	be significantly longer than the observed soliton steps in order to
	observe only the effect of the fast modulation. Now the thermal triangle
	of the resonance is moved towards the EOM modulation by changing the
	offset of the laser scan.
	When the end of the thermal triangle reaches the EOM modulation, it
	can either move over it without any effect or it can end at the modulation.
	For the first case, the EOM modulation is either not deep enough or
	not long enough (as a starting point the duration should be of the order
	of the duration of the steps, the modulation depth has to be tested).
	When the thermal triangle always ends at the modulation without any
	visible soliton steps, the EOM modulation is too strong (either too
	long or too deep or both). The desired case is in between where the
	end of the triangle gets ``stuck'' at the EOM modulation and where
	the soliton steps go over and beyond the EOM modulation (Fig.\ref{fig:powerKick}c).
	This case should also be stable with respect to small perturbations
	of the scan offset of the laser.
	
	Now the slow modulation offset can be reduced such that the power
	increase starts within the soliton step after the EOM modulation.
	The increased power stabilizes the soliton to a certain extent and the
	steps become longer (Fig.\ref{fig:powerKick}d) until, for the right
	parameters, it can be obtained as a stable, steady state. If a very fast
	increase of power with the AOM perturbs the system too much, a lower ``slow modulation speed'' can give more reliable results.
	
	With manually triggered laser scans one can monitor the traces on an oscilloscope, adjusting
	the parameters, in particular the sweep offset, and stopping at the
	time when a soliton is in a steady state. As the exact soliton state depends on the modulation parameter, one can optimize the parameters to yield a particular state. When the modulation parameters are dialed in, they usually work quite
	reliable for at least this resonance. 
	
	\begin{figure}[h!]
		\centering \includegraphics[width=1\columnwidth]{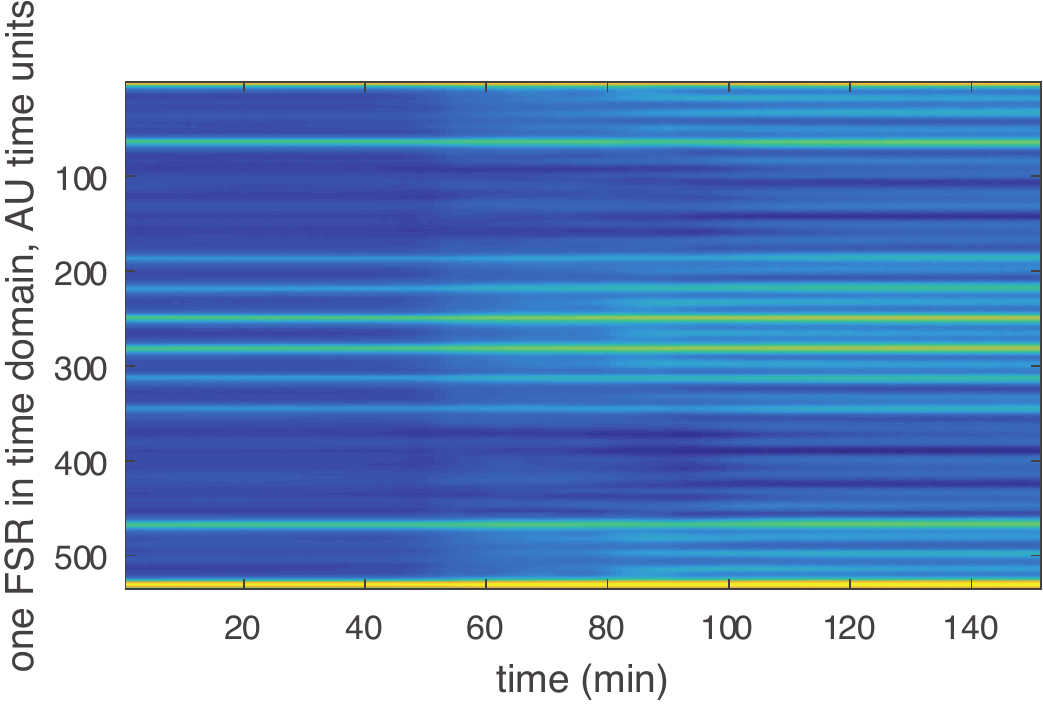}
		\caption{ \textbf{Passive long term stability of steady state soliton states in a silicon nitride microresonator.} Fourier transform of the optical spectra as shown in Fig. \ref{fig:steps}\textbf{c} cropped to one free spectral range and shown over time. Every minute one optical spectrum is recorded. The straight horizontal lines show that the solitons do not move with respect to each other over a time of more than two hours which indicates that the soliton state is passively stable without further measures. The changes in the color are due to a drifts in coupling of light to and from the photonic chip with the microresonator and resulting power fluctuations in the optical spectrum.}
		\label{fig:longTermStab} 
	\end{figure}
	
	Once a steady, stable soliton state is generated inside the microresonator it can exist for hours given that perturbations from outside (mostly drifts in pump power and pump laser frequency) are not too large (Fig.\ref{fig:longTermStab}). Small perturbations of the pump power and pump laser frequency can even be used to fully phase-stabilize the resulting Kerr frequency comb \cite{Brasch2016, Jost2015b}. 
	
	In summary we have presented a novel technique to bring soliton states in microresonators into a stable steady state via a ``power kicking'' approach. This makes the resulting pulses and coherent frequency combs usable for experiments and applications alike. Future experiments can build on these results in order to generate a particular soliton state on-demand and in a reliable way which is required for most out-of-the-lab applications as they are envisioned for integrated Kerr frequency comb sources.

	\section*{Funding Information}
	We gratefully acknowledge funding via Defense Sciences Office (DSO), DARPA (W911NF-11-1-0202); European Space Agency (ESA) (ESTEC CN 4000105962/12/NL/PA); Swiss National Science Foundation (SNSF) (Schweizerischer Nationalfonds zur F\"orderung der Wissenschaftlichen Forschung (SNF)). M.G. acknowledges support from the EPFL fellowship program co-funded by Marie Curie, FP7 Grant agreement no. 291771.
	\vspace{-1.5mm}
	
	\section*{Acknowledgments}
	The Si$_3$N$_4$ microresonator samples were fabricated in the EPFL center of MicroNanoTechnology (CMi).
	\vspace{-1.5mm}

	\bibliography{bibliography}
	

\end{document}